\documentclass[11pt]{article}
\usepackage[T1]{fontenc}
\usepackage{lmodern}
\usepackage{microtype}
\usepackage{amsmath,amssymb,amsthm,mathtools}
\usepackage{xcolor}
\usepackage[title]{appendix}
\usepackage{booktabs,array,longtable}
\usepackage[margin=0.88in]{geometry}
\usepackage[colorlinks=true,linkcolor=blue!45!black,citecolor=blue!45!black,urlcolor=blue!45!black]{hyperref}
\usepackage[nameinlink,noabbrev]{cleveref}
\crefname{equation}{eq.}{eqs.}
\Crefname{equation}{Eq.}{Eqs.}

\newcommand{\ii}{\mathrm{i}}

\newcommand{\pb}[2]{\left\{#1,#2\right\}}
\newcommand{\dd}{\mathrm{d}}

\title{Superintegrability of a resonant ghostly tri-Hamiltonian model:\\
	polynomial first integrals and transcendental completion}
\author{Andreas Fring$^{1}$ and Ian Marquette$^{2}$\\[0.5ex]
	\small $^{1}$Department of Mathematics, City St George's, University of London,\\
	\small Northampton Square, London EC1V 0HB, UK\\
	\small $^{2}$Department of Mathematical and Physical Sciences, La Trobe University,\\
	\small Bendigo VIC 3552, Australia\\[0.5ex]
	\small E-mail: \href{mailto:a.fring@city.ac.uk}{\texttt{a.fring@city.ac.uk}},
	\small \href{mailto:i.marquette@latrobe.edu.au}{\texttt{i.marquette@latrobe.edu.au}}}
\date{}

\hypersetup{pdftitle={Superintegrability of the resonant ghostly three-dimensional model},pdfauthor={Andreas Fring and Ian Marquette}}
\begin{document}
	\maketitle
	
	\begin{abstract}
We study classical and quantum superintegrability of a resonant three-dimensional Hamiltonian with indefinite kinetic energy and three Poisson descriptions of the same flow.  Besides three commuting quadratic Hamiltonians, the polynomial first-integral ring contains two primitive degree-four generators, $I_4^{\rm o}$ and $I_4^{\rm e}$, of odd and even momentum parity.  A single octic relation leaves the polynomial invariant ring with transcendence degree four.  Solving the complete first-integral equation shows that every rational first integral is a rational function of $H_1,H_2,H_3,I_4^{\rm o}$, so that the rational invariant field likewise has generic functional rank four.  A branch-free transcendental invariant raises the generic smooth functional rank to five on the invariant open set $\Delta>0$. The system is therefore minimally superintegrable in the polynomial and rational classes and maximally superintegrable on this invariant domain when transcendental integrals are admitted.  Weyl quantisation yields exact higher-order differential symmetries in all three Poisson realisations.  The branch-free fifth classical invariant also admits a densely defined nonlocal quantum counterpart, providing a concrete quantum realisation of the transcendental completion.
	\end{abstract}

	\section{Introduction}
	\label{sec:introduction}
	
Superintegrability concerns Hamiltonian systems possessing more functionally independent first integrals than are required for Liouville integrability.  Systems now termed superintegrable were already studied in the context of higher symmetries in quantum mechanics \cite{frivs1965higher}.  The general mathematical framework was introduced by Nekhoroshev under the name degenerate integrability \cite{nekhoroshev1972action}, while the modern term superintegrability appears in the work of Wojciechowski \cite{wojciechowski1983}; see also the classical analysis of Evans \cite{evans1990} and the review \cite{MillerPostWinternitz2013}. In general, for $n$ degrees of freedom, Liouville integrability requires $n$ independent integrals in involution, including the Hamiltonian, whereas a superintegrable system possesses more than $n$, with the maximal number being $2n-1$.  Thus, for three degrees of freedom, a fourth independent integral establishes superintegrability and a fifth makes the system maximally superintegrable.  The Kepler problem and the isotropic harmonic oscillator are standard examples, in which additional constants of motion are closely related to dynamical symmetry and spectral degeneracy \cite{evans1990,MillerPostWinternitz2013}.  For commensurable anisotropic frequencies, the additional integrals may have higher momentum order \cite{EvansVerrier2008,Ranada2010,Marquette2011}. For nondegenerate Pais-Uhlenbeck oscillators \cite{pais1950field} with commensurable frequencies, maximal superintegrability has been established through their relation to decoupled oscillators \cite{andrzejewski2014conf}.  The fully degenerate equal-frequency case may be semisimple or may possess a nontrivial Jordan structure. Here we study the latter case.
	
Our starting point is the three-dimensional ghostly model introduced in \cite{FringMarquette2026}, which provides a Hamiltonian realisation of the fully degenerate sixth-order Pais-Uhlenbeck oscillator, a prototypical higher time-derivative system. The term ``ghostly'' refers here to the indefinite kinetic energy.  Its classical flow contains two complex-conjugate Jordan chains of length three: the variables within each chain share the same oscillation frequency, while the non-semisimple part of the flow couples successive members and produces secular terms proportional to $t$ and $t^2$. In our analysis in \cite{FringMarquette2026} we constructed first-order intertwining operators, identified a hidden spectrum-generating $\mathfrak u(2,1)$ algebra, and exhibited three quadratic Hamiltonians paired with three constant Poisson tensors that generate the same phase-space vector field. We also found a higher-order element commuting with the complete Hamiltonian triple, but this element is polynomially reducible in terms of the quadratic Hamiltonians and therefore does not provide an additional functionally independent first integral.  The question of superintegrability consequently remained open.
	
Here, we first construct a momentum-odd polynomial integral $I_4^{\rm o}$ of total phase-space degree four and momentum order three.  It is conserved by all three paired Hamiltonian systems and is generically functionally independent of $H_1,H_2,H_3$, thereby establishing superintegrability.  The hidden $\mathfrak u(2,1)$ structure provides a compact factorisation of this integral, although only the canonical Poisson realisation carries the complete hidden Lie algebra in its original form.
	
The resonant Jordan-chain structure then allows the polynomial first-integral problem to be treated systematically.  Besides the three quadratic Hamiltonians and $I_4^{\rm o}$, a second primitive degree-four generator $I_4^{\rm e}$ of even momentum parity and momentum order four appears.  Using the locally nilpotent part of the Jordan flow and its weight decomposition, we determine the complete polynomial first-integral ring.  It is generated by $H_1,H_2,H_3,I_4^{\rm o},I_4^{\rm e}$, but these five generators obey an algebraic relation of degree eight.  Consequently the polynomial sector has transcendence degree four, and $I_4^{\rm e}$ does not supply the fifth functionally independent integral required for maximal superintegrability.
	
We next solve the complete first-integral equation without imposing a polynomial ansatz.  A set of characteristic coordinates gives the general local solution explicitly and makes it possible to determine the rational invariant field.  We prove that every rational first integral is a rational function of $H_1,H_2,H_3,I_4^{\rm o}$, so no fifth independent rational integral exists at any momentum order.  The missing fifth invariant is represented locally by a transcendental characteristic containing a logarithmic phase.  Although the logarithmic representative is branch dependent, exponentiation yields a branch-independent first integral $Z_5$ that is single-valued on $\Delta>0$.  Since $\Delta$ is itself conserved, this open set is invariant under the flow, and the transcendental sector raises the generic smooth functional rank to five there.  The system is therefore minimally superintegrable in the polynomial and rational classes, but maximally superintegrable on the invariant open set $\Delta>0$ when transcendental first integrals are admitted.  The existence of five functionally independent first integrals on such an invariant domain is consistent with the local straightening of a nonvanishing vector field; what is specific here is their explicit construction and the proof that the fifth integral cannot be chosen rationally.
	
Non-polynomial first integrals have also been considered in other settings.  Within Darboux theory, exponential factors of the form $\exp(g/h)$ may enter first integrals \cite{LlibreZhang2009,MaciejewskiPrzybylska2020}; in particular, closely related products of a polynomial factor with the exponential of a rational function occur in integrable polynomial flows \cite{MaciejewskiPrzybylska2020}.  Non-polynomial first integrals of Hamiltonian geodesic flows have been constructed in \cite{Galajinsky2021,Agapov2022}, while a superintegrable two-dimensional metric with a transcendental additional integral was recently obtained in \cite{GineSinelshchikov2024}.  The situation considered here is different: we determine the complete polynomial and rational invariant structures of a non-semisimple resonant Hamiltonian flow, prove that no rational fifth integral exists, and show that maximal superintegrability is achieved by a transcendental completion.
	
Finally, we quantise all three Poisson descriptions by realising the phase-space variables in Darboux coordinates and applying Weyl ordering.  The polynomial integrals give exact differential symmetries of orders three and four for each paired Hamiltonian, and their commutator algebra contains genuine quantum ordering corrections.  In particular, Weyl ordering a complete classical product must be distinguished from multiplying separately quantised factors.  The branch-independent classical fifth invariant also has a natural nonlocal quantum counterpart.  The first-order operator entering its inverse becomes a multiplication operator after a partial Fourier transform, which shows that the inverse is densely defined, while the resulting exponential can be defined on suitable analytic vectors.  Thus the fifth symmetry can be realised beyond a purely formal level.  A full operator-theoretic analysis, and in particular the appropriate notion of independence needed to formulate quantum maximal superintegrability, is left open.
	
The paper is organised as follows:  In section~2 we introduce the tri-Hamiltonian structure and the first additional integral $I_4^{\rm o}$.  In section~3 we relate this construction to the hidden $\mathfrak u(2,1)$ algebra.  In section~4 we develop the Jordan-chain description, construct a further integral $I_4^{\rm e}$ and determine the complete polynomial first-integral ring and its Poisson algebra.  In section~5 we solve the full first-integral equation, determine the rational invariant field and construct the transcendental fifth integral. In section~6 we develop the Weyl quantisation, the polynomial quantum symmetries and the nonlocal quantum counterpart of the fifth classical invariant.  Our conclusions are stated in section~7.  The appendices collect the remaining hidden-algebra data and the explicit third-order differential symmetry operators.

\section{Tri-Hamiltonian structure and a polynomial first integral}

We consider a dynamical system on the six-dimensional phase space
\begin{equation}
	\boldsymbol z=(x,y,z,p_x,p_y,p_z)^{\mathsf T}
\end{equation}
whose flow admits three distinct Hamiltonian descriptions.  This tri-Hamiltonian structure was originally identified for the present model in \cite{FringMarquette2026}.  More precisely, there exist three Hamiltonian functions $H_r$ and three nondegenerate Poisson tensors $J_r$, $r=1,2,3$, such that the same phase-space vector field $\boldsymbol X$ is generated by each pair,
\begin{equation}
	\boldsymbol X (\boldsymbol z) = J_1\nabla H_1 = J_2\nabla H_2 = J_3\nabla H_3 .
	\label{eq:trihamiltonian}
\end{equation}
We shall refer to \eqref{eq:trihamiltonian} as the tri-Hamiltonian structure of the model.  Related tri-Hamiltonian formulations occur, for example, in \cite{tri1,tri2,tri3,felski2026three}.

Our first aim is to determine whether this structure admits an additional first integral beyond the three quadratic Hamiltonians and thereby establishes superintegrability.
	
For a Poisson tensor $J$, we use the convention
	\begin{equation}
		\{F,G\}_{J} := (\nabla F)^{\mathsf T}J\nabla G .
		\label{eq:Poisson-bracket}
	\end{equation}
	The three Poisson tensors are
	\begin{equation}
		J_r= \begin{pmatrix} 0&M_r\\ -M_r^{\mathsf T}&0 \end{pmatrix},
		\label{Jblock}
	\end{equation}
	with
		\begin{equation}
		M_1=I_3,\qquad
		M_2= \begin{pmatrix} 2&1&1\\ -1&0&-1\\ 1&1&1 \end{pmatrix}, \qquad
		M_3= \begin{pmatrix} 2&1&0\\ -1&0&0\\ 0&0&1 \end{pmatrix}.
		\label{eq:Mmatrices-short}
	\end{equation}
	All three Poisson tensors are constant and nondegenerate, with $\det J_r=1$.  Their constancy implies the Jacobi identity, so they define symplectic Poisson structures. The corresponding Hamiltonians are
	\begin{align}
		H_1={}&p_x^2-p_y^2+p_z^2+(g^2+\lambda^2)x^2+(g^2-\lambda^2)y^2+2g^2xy
		-4g\lambda xz-4g\lambda yz+\lambda^2z^2,\label{eq:H1}\\
		H_2={}&p_x^2-p_y^2+p_z^2-2p_xp_z+2p_yp_z+(g+\lambda)^2x^2
		+(g^2+2g\lambda-\lambda^2)y^2\notag\\
		&\quad+2g(g+2\lambda)xy-2\lambda(2g+\lambda)xz
		-2\lambda(2g+\lambda)yz+\lambda^2z^2,\label{eq:H2}\\
		H_3={}&2p_xp_y-2p_y^2+p_z^2+g^2x^2+(g^2-2\lambda^2)y^2
		+(2g^2-2\lambda^2)xy\notag\\
		&\quad-4g\lambda xz-4g\lambda yz+\lambda^2z^2,\label{eq:H3}
	\end{align}
	where $g,\lambda\in\mathbb R\setminus\{0\}$ are fixed parameters. A direct calculation verifies equations (\ref{eq:trihamiltonian}), confirming that the three Hamiltonian pairs generate the same flow.
	Moreover,
	\begin{equation}
		\{H_1,H_2\}_{J_r} = \{H_1,H_3\}_{J_r} = \{H_2,H_3\}_{J_r} =0, \qquad r=1,2,3.
		\label{eq:all-involution}
	\end{equation}
	Thus $H_1,H_2,H_3$ are conserved and mutually commuting in every
	Poisson realisation.
	
	We now seek an additional first integral.  For convenience, we introduce the quadratic
	phase-space functions
\begin{align}
	A={}&(p_x-p_y)^2-\lambda^2(x+y)^2,\label{eq:A-direct}\\
	B={}&(p_x-p_y)(gz+\lambda y)-p_z\bigl[g(x+y)-\lambda z\bigr]
	-\lambda p_y(x+y),\label{eq:B-direct}\\
	C={}&p_z^2-2\lambda(x+y)(gz+\lambda y)-2p_y(p_x-p_y)
	-\bigl[g(x+y)-\lambda z\bigr]^2,\label{eq:C-direct}\\
	D={}&(x+y)(p_x-p_y),\label{eq:D-direct}
\end{align}
	and define
	\begin{equation}
		I_4^{\rm o}=128g^6(AB-\lambda DC).
		\label{eq:Icompact}
	\end{equation}	
The integral $I_4^{\rm o}$ has total phase-space degree four and momentum order three. The superscript ${\rm o}$ refers to its odd momentum parity, $I_4^{\rm o}(q,-p) = -I_4^{\rm o}(q,p) $.
	
To write the Poisson brackets of $I_4^{\rm o}$ compactly, we introduce the dependent combinations
	\begin{equation}
		\Delta:=H_1-H_3
		=(p_x-p_y)^2+\lambda^2(x+y)^2,    \quad \text{and} \quad	K:=256g^8\Delta^2 .
	\end{equation}
Both are functions of the three quadratic Hamiltonians and therefore do not constitute additional independent first integrals.  Notice also that $\Delta\geq0$ on the real phase space, which we will exploit below.

The brackets of  $I_4^{\rm o}$ with the three Hamiltonians are summarised in \cref{tab:classical-table}.  Every entry is the displayed coefficient times $K$.
	\begin{table}[ht]
		\centering
		\caption{Classical brackets $\pb{H_a}{I_4^{\rm o}}_{J_r}/K$.}
		\label{tab:classical-table}
		\begin{tabular}{c|ccc}
			\toprule
			& $H_1$ & $H_2$ & $H_3$\\
			\midrule
			$J_1$ & $0$ & $1/g$ & $-\lambda/g^2$\\
			$J_2$ & $(\lambda-g)/g^2$ & $0$ & $-1/g$\\
			$J_3$ & $\lambda/g^2$ & $(g+\lambda)/g^2$ & $0$\\
			\bottomrule
		\end{tabular}
	\end{table}
	
The diagonal entries vanish, so $I_4^{\rm o}$ is conserved by each paired system $(H_r,J_r)$.  The off-diagonal entries describe its brackets with Hamiltonians that do not generate the corresponding paired flow. It then follows directly from the table that
	\begin{equation}
		\pb{\Delta}{I_4^{\rm o}}_{J_r}=\frac{\lambda}{g^2}K, 
		\qquad r=1,2,3.
		\label{eq:DeltaIclassical}
	\end{equation}
Since $K=256g^8\Delta^2 $, the Poisson algebra generated by $H_1,H_2,H_3,I_4^{\rm o}$ closes polynomially.
	
The three Hamiltonians are generically independent because
	\begin{equation}
		\det\frac{\partial(H_1,H_2,H_3)}{\partial(p_x,p_y,p_z)}=8(p_x-p_y)^3.
		\label{eq:three-rank}
	\end{equation}
Moreover, a direct four-by-four Jacobian minor gives the explicit result
	\begin{equation}
		\det\frac{\partial(H_1,H_2,H_3,I_4^{\rm o})}
		{\partial(z,p_x,p_y,p_z)}
		=-1024g^6(p_x-p_y)\bigl[g(p_x-p_y)+\lambda p_z\bigr]
		\, \bigl[(p_x-p_y)^2+\lambda^2(x+y)^2\bigr]^2.
		\label{eq:four-rank}
	\end{equation}
Hence $H_1,H_2,H_3,I_4^{\rm o}$ are functionally independent on a generic open subset of phase space, and therefore
\begin{equation}
	\operatorname{rank}\bigl(dH_1,dH_2,dH_3,dI_4^{\rm o}\bigr)=4
\end{equation}
generically.  The additional integral $I_4^{\rm o}$ therefore establishes superintegrability of each paired Hamiltonian system. Whether a fifth independent first integral exists, and in which function class, will be addressed below.

\section{Hidden $\mathfrak u(2,1)$ interpretation of $	I_4^{\rm o}$}
\label{sec:hidden-form}

The canonical Hamiltonian realisation $(H_1,J_1)$ was shown in \cite{FringMarquette2026} to possess a hidden spectrum-generating $\mathfrak u(2,1)$ algebra.  The individual generators of this algebra are not, in general, first integrals.  Its relevance here is instead that the three quadratic Hamiltonians and the additional polynomial integral $I_4^{\rm o}$ admit particularly simple expressions in terms of its classical phase-space symbols.

Only four noncentral symbols are required for the present discussion:
\begin{align}
	h_{\mathrm{cl}}
	&= 2\ii\left[ \frac{g}{\lambda}z(p_x-p_y) +yp_x+xp_y \right], \label{eq:hcl-short} \\ e_{\mathrm{cl}}
	&= 2\ii\left[ -\frac{g}{\lambda}(x+y)(p_x-p_y) -p_z(x+y) +z(p_x-p_y) \right], \label{eq:ecl-short} \\ (v_1)_{\mathrm{cl}}
	&= -2\frac{g}{\lambda}(p_x-p_y)^2 +2p_z(p_y-p_x) -2\lambda^2z(x+y), \label{eq:v1cl-short} \\ (v_2)_{\mathrm{cl}}
	&= -\lambda^2(x+y)^2-(p_x-p_y)^2. \label{eq:v2cl-short}
\end{align}
The subscript ``cl'' denotes the corresponding commuting classical phase-space symbol.

To distinguish the central quadratic generator of the hidden algebra from the quartic closure element introduced in section~2, we denote the former by $K_{\mathrm{hid}}$.  Its classical symbol is
\begin{equation}
	(K_{\mathrm{hid}})_{\mathrm{cl}} = (v_1)_{\mathrm{cl}} -\frac{g}{\lambda}(v_2)_{\mathrm{cl}} -\frac{\lambda}{g}H_1.
	\label{eq:Khid-short}
\end{equation}
With this normalisation, the three Hamiltonians take the form
\begin{align}
	H_1 &= \frac{g}{\lambda}(v_1)_{\mathrm{cl}} -\frac{g^2}{\lambda^2}(v_2)_{\mathrm{cl}} -\frac{g}{\lambda}(K_{\mathrm{hid}})_{\mathrm{cl}}, \notag \\
	H_2 &= \left(1+\frac{g}{\lambda}\right)(v_1)_{\mathrm{cl}} - \left( \frac{g^2}{\lambda^2} +2\frac{g}{\lambda} \right)(v_2)_{\mathrm{cl}} -\frac{g}{\lambda}(K_{\mathrm{hid}})_{\mathrm{cl}}, \label{eq:Hhidden-short} \\
	H_3 &= \frac{g}{\lambda}(v_1)_{\mathrm{cl}} + \left( 1-\frac{g^2}{\lambda^2} \right)(v_2)_{\mathrm{cl}} -\frac{g}{\lambda}(K_{\mathrm{hid}})_{\mathrm{cl}}.
	\notag
\end{align}
Most importantly, the momentum-odd quartic integral found in section~2 factorises as
\begin{equation}
	I_4^{\rm o} = 64\ii g^6\lambda \left[ h_{\mathrm{cl}}(v_2)_{\mathrm{cl}} +\frac12 e_{\mathrm{cl}}(v_1)_{\mathrm{cl}} -\frac{g}{\lambda}e_{\mathrm{cl}}(v_2)_{\mathrm{cl}} \right].
	\label{eq:I4hiddencl-short}
\end{equation}
Thus, in the canonical realisation, the additional polynomial first integral is naturally embedded in the same hidden-algebraic structure that organises the quadratic Hamiltonians.

For the canonical Poisson tensor $J_1$, the symbols above form part of the genuine hidden $\mathfrak u(2,1)$ realisation of \cite{FringMarquette2026}.  If the same phase-space functions are equipped instead with the Poisson brackets defined by $J_2$ or $J_3$, they need not satisfy the same Lie-algebra relations.  This distinction does not affect \eqref{eq:Hhidden-short} and \eqref{eq:I4hiddencl-short}, which are algebraic identities between phase-space functions and are therefore independent of the choice of Poisson tensor.  
Additional hidden-algebra data, including its $\mathfrak{sl}_2$ structure and the relation between the noncanonical brackets and algebra-preserving pullbacks, are collected in Appendix~\ref{app:hidden-symbols}.

\section{Jordan-chain structure and polynomial first integrals}
	\label{sec:second-quartic-classical}
	
The polynomial first integrals can be classified systematically by exploiting the two complex-conjugate Jordan chains of length three underlying the resonant flow, identified for the present model in \cite{FringMarquette2026}.  Their semisimple parts carry eigenvalues $\pm2\ii\lambda$, while the nilpotent part couples successive members within each chain. Denoting with  $\boldsymbol X$ the common Hamiltonian vector field defined in equation \eqref{eq:trihamiltonian}, and definind
	\begin{equation}
		\mathcal X:=\boldsymbol X\cdot\nabla
		=\{\,\cdot\,,H_r\}_{J_r},
		\qquad r=1,2,3,
		\label{eq:common-derivation}
	\end{equation}
to be the corresponding derivation on phase-space functions.  Then a function $F$ is a first integral precisely when $\mathcal XF=0$.

The two Jordan chains may be represented by the complex linear phase-space functions
	\begin{equation}
		\begin{aligned}
			u_0&=\ii(p_x-p_y)-\lambda(x+y),\\
			u_1&=-\frac{\ii p_z+g(x+y)-\lambda z}{2g},\\
			u_2&=\frac{\ii(p_x+p_y)-\lambda(x-y)+2gz}{8g^2},
		\end{aligned}
		\qquad
		\begin{aligned}
			v_0&=\ii(p_x-p_y)+\lambda(x+y),\\
			v_1&=\frac{\ii p_z-g(x+y)+\lambda z}{2g},\\
			v_2&=\frac{\ii(p_x+p_y)+\lambda(x-y)-2gz}{8g^2}.
		\end{aligned}
		\label{eq:uvchains-short}
	\end{equation}
These are scaled versions of the first-order intertwining symbols introduced in \cite{FringMarquette2026}. We note that the symbols $v_j$ used here are chain variables and should not be confused with the quadratic hidden-algebra symbols $(v_m)_{\rm cl}$ of the preceding section.
	
Their evolution under the common flow is
	\begin{align}
		\mathcal Xu_0&=2\ii\lambda u_0,
		&
		\mathcal Xu_1&=2\ii\lambda u_1+\ii u_0,
		&
		\mathcal Xu_2&=2\ii\lambda u_2+\ii u_1,
		\notag\\
		\mathcal Xv_0&=-2\ii\lambda v_0,
		&
		\mathcal Xv_1&=-2\ii\lambda v_1+\ii v_0,
		&
		\mathcal Xv_2&=-2\ii\lambda v_2+\ii v_1 .
			\label{eq:chain-evolution-short}
	\end{align}
It is useful to rewrite the derivation in the form
	\begin{equation}
		-\ii\mathcal X=2\lambda\mathcal W+\mathcal N, \label{eq:XWN}
	\end{equation}
where
	\begin{equation}
		\mathcal W = \sum_{j=0}^2u_j\partial_{u_j} - \sum_{j=0}^2v_j\partial_{v_j}, \qquad \mathcal N = u_0\partial_{u_1}+u_1\partial_{u_2} +v_0\partial_{v_1}+v_1\partial_{v_2}.
		\label{eq:WN-main-j22}
	\end{equation}
	
This is the Jordan-Chevalley decomposition of the derivation $-i\mathcal X$: $2\lambda\mathcal W$ is its semisimple part and $\mathcal N$ its locally nilpotent part. Moreover, we observe
\begin{equation}
	[\mathcal W,\mathcal N]=0.  \label{WNcomm}
\end{equation}
For a semisimple derivation, a polynomial $F$ is said to have weight $m$ if $\mathcal WF=mF$. The polynomial algebra therefore decomposes into weight spaces
\begin{equation}
	\mathcal A=\bigoplus_m\mathcal A_m,
	\qquad
	\mathcal A_m=\{F:\mathcal WF=mF\}.
\end{equation}
Since $\mathcal W$ is a derivation, weights are additive under	multiplication.
	
In our case we have
	\begin{equation}
		\mathcal Wu_j=u_j,\qquad \mathcal Wv_j=-v_j,
	\end{equation}
so that
	\begin{equation}
		w(u_j)=1,\qquad w(v_j)=-1, \qquad j=0,1,2.
	\end{equation}
Thus the weight of a monomial is the number of $u$ factors minus the number of $v$ factors.  The nilpotent derivation $\mathcal N$, by contrast, does not define a grading: it moves variables along the Jordan chains,
	\begin{equation}
	\mathcal Nu_0=0,\qquad
	\mathcal Nu_1=u_0,\qquad
	\mathcal Nu_2=u_1,
\end{equation}
and similarly for the $v_j$.  Since $\mathcal W$ and $\mathcal N$ commute, (\ref{WNcomm}), the nilpotent derivation $\mathcal N$ maps each $\mathcal W$-eigenspace into itself and therefore preserves the weight decomposition.

We now decompose an arbitrary polynomial $\Psi$ into its $\mathcal W$-weight components,
	\begin{equation}
		\Psi=\sum_m\Psi_m .
	\end{equation}
Since $\mathcal N$ preserves both total degree and weight, the first-integral condition $\mathcal X\Psi=0$ gives, separately in each weight space,
	\begin{equation}
		(2\lambda m+\mathcal N)\Psi_m=0.
	\end{equation}
For $m\neq0$, the operator $2\lambda m+\mathcal N$ is invertible on each finite-dimensional homogeneous component because $\mathcal N$ is nilpotent there.  Hence $\Psi_m=0$ for all $m\neq0$.  For $m=0$ the condition reduces to
	\begin{equation}
		\mathcal N\Psi_0=0.
	\end{equation}
Consequently, the polynomial first integrals are precisely the weight-zero elements of $\ker\mathcal N$.
	
At total phase-space degree two, every weight-zero polynomial has the form
	\begin{equation}
		F_2=\sum_{i,j=0}^{2}c_{ij}u_iv_j .
	\end{equation}
Solving $\mathcal NF_2=0$ gives a three-dimensional space with basis
	\begin{equation}
		q_0=u_0v_0,\qquad q_1=u_0v_1-u_1v_0,\qquad q_2=u_0v_2-u_1v_1+u_2v_0 .
		\label{eq:q012-short}
	\end{equation}
In terms of the original Hamiltonians these are
	\begin{equation}
		q_0=H_3-H_1,\qquad q_1=\frac{H_2-H_1}{2g},\qquad q_2=-\frac{H_1}{4g^2}.
		\label{eq:qH-short}
	\end{equation}
Thus the three quadratic polynomial first integrals obtained from the Jordan chains are precisely the three Hamiltonians already encountered in section 2.
	
To determine the next homogeneous sector, we first introduce the $\mathcal N$-invariant combinations
	\begin{equation}
		U:=2u_0u_2-u_1^2, \qquad V:=2v_0v_2-v_1^2. \label{eq:UV-short}
	\end{equation}
They have weights
	\begin{equation}
		w(U)=2,\qquad w(V)=-2,
	\end{equation}
and hence the combinations
	\begin{equation}
		P:=v_0^2U,\qquad Q:=u_0^2V,\qquad R:=UV
		\label{eq:PQR-short}
	\end{equation}
have weight zero and lie in $\ker\mathcal N$.  Moreover, they satisfy the identities
	\begin{equation}
		P+Q=2q_0q_2-q_1^2, \qquad PQ=q_0^2R. \label{eq:PQrelations-short}
	\end{equation}
	
These expressions also give a direct completeness result at degree four.  A general weight-zero homogeneous polynomial of degree four contains two $u$ factors and two $v$ factors.  A direct linear-algebra calculation of $\mathcal NF_4=0$ in this finite-dimensional space gives
	\begin{equation}
		\dim\mathcal I_4=8,
	\end{equation}
where $\mathcal I_4$ denotes the space of homogeneous degree-four polynomial first integrals.  A convenient basis is
	\begin{equation}
		\mathcal I_4= \operatorname{span} \left\{ q_0^2,\, q_0q_1,\, q_0q_2,\, q_1^2,\, q_1q_2,\, q_2^2,\, P-Q,\, R \right\}.
		\label{eq:I4-complete}
	\end{equation}
The first six elements are products of the quadratic first integrals. The quotient by this reducible sector is therefore two-dimensional,
	\begin{equation}
		\dim \frac{\mathcal I_4} {\operatorname{span}\{q_iq_j:0\leq i\leq j\leq2\}} =2.
		\label{eq:primitive-dim}
	\end{equation}
Hence there are precisely two primitive polynomial directions at total phase-space degree four.
	
The first is the momentum-odd integral constructed in section 2, which in the Jordan variables takes the particularly simple form
	\begin{equation}
		I_4^{\rm o} =128\ii g^8(P-Q). \label{eq:I4o-PQ}
	\end{equation}
The second primitive direction may be chosen momentum even.  We define
	\begin{equation}
		I_4^{\rm e}:=16g^4R=F^2+G^2, \label{eq:J4definition-short}
	\end{equation}
where
	\begin{align}
		F={}&p_x^2-p_y^2-p_z^2 +g^2(x+y)^2-\lambda^2x^2+\lambda^2y^2+\lambda^2z^2, \label{eq:F-short}\\
		G={}&2\bigl[ -gz(p_x-p_y)+gp_z(x+y) +\lambda(xp_x+yp_y-zp_z) \bigr]. \label{eq:G-short}
	\end{align}
Both $I_4^{\rm o}$ and $I_4^{\rm e}$ are homogeneous of total phase-space degree four.  The former is odd and the latter even under $p\mapsto-p$, with momentum orders three and four, respectively. Since $R$ is a weight-zero element of $\ker\mathcal N$, the characterisation above immediately implies $\mathcal X R=0$, so that $I_4^{\rm e}$ is conserved by each paired Hamiltonian system
	\begin{equation}
		\{H_r,I_4^{\rm e}\}_{J_r}=0, \qquad r=1,2,3. \label{eq:J4paired-short}
	\end{equation}
	
The preceding calculation establishes completeness at total phase-space degree four. 
To determine the polynomial first-integral ring at all degrees, we
now use the invariant theory of locally nilpotent derivations.
Theorem~2.1 of Wehlau \cite{Wehlau2014} considers the
Weitzenb\"ock derivation
\begin{equation}
\Delta=\sum_{i=1}^{n} \left( x_i\frac{\partial}{\partial y_i} +y_i\frac{\partial}{\partial z_i} \right)
\end{equation}
and gives a minimal generating set for its kernel when all Jordan
blocks have length three.  Our derivation $\mathcal N$ is precisely
the case $n=2$, under the identification
\begin{equation}
(x_1,y_1,z_1)=(u_0,u_1,u_2),
\qquad
(x_2,y_2,z_2)=(v_0,v_1,v_2).
\end{equation}
Indeed, Wehlau's generators mentioned in (1), (2), (3) of the theorem are in this case to
\begin{equation}
x_1=u_0,\,\, x_2=v_0, \,\,  x_1y_2-x_2y_1=q_1, \,\, 
2x_1z_1-y_1^2=U,\,\, 
x_1z_2-y_1y_2+z_1x_2=q_2,\,\, 
2x_2z_2-y_2^2=V.
\end{equation}
There are no determinant-type generators as in (4) of the theorem, since only two Jordan
chains are present.  Hence
\begin{equation}
	\ker\mathcal N
	=
	\mathbb C[u_0,v_0,q_1,q_2,U,V].
	\label{eq:Wehlau-generators}
\end{equation}
with $\mathcal W$-weights $(1,-1,0,0,2,-2)$, respectively.  Indeed, a weight-zero monomial $u_0^a v_0^b U^cV^d$ satisfies $a-b+2c-2d=0$ and can therefore be factorised into powers of $u_0v_0$, $v_0^2U$, $u_0^2V$ and $UV$. The weight-zero subring is consequently generated by
	\begin{equation}
		q_0=u_0v_0,\qquad q_1,\qquad q_2,\qquad P=v_0^2U,\qquad Q=u_0^2V,\qquad R=UV .
		\label{weightzero}
	\end{equation}
	The first relation in \eqref{eq:PQrelations-short} eliminates $Q$,
	while the second then becomes a relation among the remaining five
	generators. Thus the complete polynomial
	first-integral ring is generated by
	\begin{equation}
		q_0,\qquad q_1,\qquad q_2,\qquad P-Q,\qquad R,
	\end{equation}
	or, equivalently, by
	\begin{equation}
		H_1,\qquad H_2,\qquad H_3,\qquad I_4^{\rm o},\qquad I_4^{\rm e}. \label{eq:five-poly-generators}
	\end{equation}
In particular, no new primitive polynomial generators occur at higher degree.
	
The algebraic dependence between the two degree-four generators is most conveniently expressed in the original variables.  With
	\begin{equation}
		\Delta:=H_1-H_3, \qquad \Theta:=2H_1\Delta-(H_2-H_1)^2, \label{eq:Theta-short}
	\end{equation}
equation \eqref{eq:PQrelations-short} gives
	\begin{equation}
		4\Delta^2 I_4^{\rm e} = \Theta^2 + \left(\frac{I_4^{\rm o}}{32g^6}\right)^2 . \label{eq:octicrelation-short}
	\end{equation}
Hence $I_4^{\rm e}$ is polynomially primitive but not functionally independent of $H_1,H_2,H_3,I_4^{\rm o}$.  On the generic chart $\Delta\neq0$ it is already a rational function of those four integrals,
	\begin{equation}
		I_4^{\rm e} = \frac{\Theta^2+ (I_4^{\rm o})^2/(1024g^{12})} {4\Delta^2}.
		\label{eq:I4e-rational}
	\end{equation} 
On the real phase space, the chart $\Delta\neq0$ is precisely the open set $\Delta>0$, which will play a central role in section~5. Since we established in section 2 that $H_1,H_2,H_3,I_4^{\rm o}$ have generic functional rank four, it follows that
	\begin{equation}
		\operatorname{rank}\,
		\dd(H_1,H_2,H_3,I_4^{\rm o},I_4^{\rm e})=4
		\label{eq:rank-five-polynomial}
	\end{equation}
generically.
	
Since $q_0,q_1,q_2,P-Q$ are algebraically independent on a generic open set, as follows equivalently from the rank-four result for $H_1,H_2,H_3,I_4^{\rm o}$ in \eqref{eq:four-rank}, the polynomial first-integral ring has transcendence degree four.  The kernel of the surjective homomorphism from the polynomial algebra in the five generators is therefore a prime ideal of height one.  Since the polynomial ring in five abstract generators is a unique factorisation domain, this ideal is principal.  Moreover, the relation \eqref{eq:octicrelation-short}, regarded as a polynomial in $I_4^{\rm e}$, is primitive and linear; its coefficient $4\Delta^2$ is coprime to its constant term.  It is therefore irreducible and generates the kernel. Thus, writing $\mathcal A_{\rm pol}$ for the algebra of polynomial first integrals over $\mathbb C$, the preceding results give the presentation
	\begin{equation}
		\mathcal A_{\rm pol} \simeq \frac{ \mathbb C[ H_1,H_2,H_3,I_4^{\rm o},I_4^{\rm e}] } { \left\langle 4(H_1-H_3)^2I_4^{\rm e} -\Theta^2 -(I_4^{\rm o})^2/(1024g^{12}) \right\rangle }.
		\label{eq:ringquotient-j22}
	\end{equation}
Thus the polynomial first-integral ring has five generators but transcendence degree four.  The fifth functionally independent first integral required for maximal superintegrability, if it exists, must therefore lie outside the polynomial class.
	
For completeness, the Poisson brackets of the five polynomial generators also close.  With $c_{ra}$ denoting the coefficients in table \ref{tab:classical-table}, we obtain
	\begin{equation}
		\{H_a,I_4^{\rm o}\}_{J_r}=c_{ra}K, \qquad \text{and} \qquad \{H_a,I_4^{\rm e}\}_{J_r} = c_{ra}\frac{I_4^{\rm o}}{8g^4}.
		\label{eq:HJ4tablecompact-short}
	\end{equation}
Moreover we find
	\begin{equation}
		\{I_4^{\rm o},I_4^{\rm e}\}_{J_r} = 256g^6 \left( 2\lambda\Delta I_4^{\rm e} -\Theta L_r \right), \qquad r=1,2,3
		\label{eq:I4J4classical-short}
	\end{equation}
	where we abbreviated
	\begin{align}
		L_1&=(g+\lambda)H_1-gH_2,\notag\\
		L_2&=\lambda H_1+(\lambda-g)H_2+(g-\lambda)H_3,
		\label{eq:Lr-short}\\
		L_3&=(g+2\lambda)H_1-gH_2-\lambda H_3.
		\notag
	\end{align}
	
Together with equation \eqref{eq:octicrelation-short} and the brackets of section 2, these relations close the polynomial Poisson algebra.
	
\section{Complete first-integral equation and transcendental completion}
	\label{sec:pde-completion}
	
Having classified the polynomial first integrals, we now solve the first-integral equation of the common flow without imposing a polynomial or rational ansatz.  In the original phase-space coordinates, the derivation $\mathcal X$ introduced in \eqref{eq:common-derivation} is
	\begin{equation}
		\begin{aligned}
			\mathcal X={}&2p_x\partial_x-2p_y\partial_y+2p_z\partial_z
			+\left[-2(g^2+\lambda^2)x-2g^2y+4g\lambda z\right]\partial_{p_x}
			+\left[-2g^2x-2(g^2-\lambda^2)y+4g\lambda z\right]\partial_{p_y}\\
			&+\left[4g\lambda(x+y)-2\lambda^2z\right]\partial_{p_z}.
		\end{aligned}
		\label{eq:Xexplicit-short}
	\end{equation}
As discussed in the previous section, the complete first-integral problem is the first-order equation $ \mathcal XF=0$. The Jordan-chain variables introduced in \eqref{eq:uvchains-short} provide coordinates adapted to this equation.  On the chart $u_0v_0\neq0$, we set
	\begin{equation}
		s=\frac{u_1}{u_0}, \qquad t=\frac{v_1}{v_0}. \label{eq:st-def}
	\end{equation}
The evolution equations \eqref{eq:chain-evolution-short} give
	\begin{equation}
		\mathcal Xs=\mathcal Xt=\ii.
		\label{eq:st-flow}
	\end{equation}
Thus $s$ and $t$ both vary linearly along the flow, while their
difference is invariant.  Moreover, with
\begin{equation}
\mathcal X\!\left(\frac{u_2}{u_0}\right)=\ii s,
\qquad
\mathcal X\!\left(\frac{v_2}{v_0}\right)=\ii t,
\qquad
\mathcal X\!\left(\frac{s^2}{2}\right)=\ii s,
\qquad
\mathcal X\!\left(\frac{t^2}{2}\right)=\ii t,
\end{equation}
it follows that the combinations
\begin{equation}
\rho:=u_0v_0,\qquad
\delta:=s-t,\qquad
\alpha:=\frac{u_2}{u_0}-\frac{s^2}{2},
\qquad
\beta:=\frac{v_2}{v_0}-\frac{t^2}{2}
\end{equation}
are invariant under $\mathcal X$. Furthermore, since the first relation in \eqref{eq:chain-evolution-short}  implies $\mathcal X\ln u_0=2\ii\lambda$, it follows with \eqref{eq:st-flow}  that
	\begin{equation}
		\zeta=\ln u_0-2\lambda s.
		\label{eq:chi-def}
	\end{equation}
is also invariant. Thus we have
	\begin{equation}
		\mathcal X\rho = \mathcal X\delta = \mathcal X\alpha = \mathcal X\beta = \mathcal X\zeta =0.
		\label{eq:five-characteristics}
	\end{equation}
These five quantities form a complete set of local characteristics. Indeed, 
	the transformation from $ (u_0,u_1,u_2,v_0,v_1,v_2)$ to $(\rho,\delta,\alpha,\beta,\zeta,s) $ is locally invertible on $u_0v_0\neq0$, with inverse
		\begin{equation}
		\begin{aligned}
			u_0&=\exp(\zeta+2\lambda s),
			&
			u_1&=s u_0,
			&
			u_2&=\left(\alpha+\frac{s^2}{2}\right)u_0,
			\\
			v_0&=\rho\exp(-\zeta-2\lambda s),
			&
			v_1&=(s-\delta)v_0,
			&
			v_2&=\left(\beta+\frac{(s-\delta)^2}{2}\right)v_0.
		\end{aligned}
		\label{eq:inverse-char-j22}
	\end{equation}
Consequently, in these coordinates $\mathcal X=\ii\partial_s$, and the general local holomorphic solution is
	\begin{equation}
		F=\Phi(\rho,\delta,\alpha,\beta,\zeta),
		\label{eq:complete-PDE-short}
	\end{equation}
where $\Phi$ is an arbitrary local holomorphic function.  A branch of the logarithm is understood in the definition of $\zeta$.  Thus \eqref{eq:complete-PDE-short} is the complete local characteristic solution rather than an ansatz for a restricted class of first integrals.

We next determine which part of this solution is rational.  The first four characteristics are rational functions of the polynomial integrals found above.  Using
	\begin{equation}
		\Delta=H_1-H_3, \qquad \Theta=2H_1\Delta-(H_2-H_1)^2, \label{eq:DeltaTheta-repeat}
	\end{equation}
	we obtain
	\begin{equation}
		\rho=H_3-H_1, \qquad \delta=\frac{H_2-H_1}{2g\Delta}, \qquad \alpha+\beta=\frac{\Theta}{8g^2\Delta^2}, \qquad \alpha-\beta= \frac{I_4^{\rm o}}{256\ii g^8\Delta^2}.
		\label{eq:char-to-integrals-short}
	\end{equation}
	Conversely, the polynomial integrals can be recovered rationally from
	these four characteristics:
	\begin{equation}
		\begin{aligned}
			H_1&=-4g^2\rho \left(\alpha+\beta+\frac{\delta^2}{2}\right), \\
			H_2&=H_1-2g\rho\delta, \\
			H_3&=H_1+\rho, \\
			I_4^{\rm o}&= 256\ii g^8\rho^2(\alpha-\beta).
		\end{aligned}
		\label{eq:integrals-to-char}
	\end{equation}
	Hence
	\begin{equation}
		\mathbb C(\rho,\delta,\alpha,\beta) = \mathbb C(H_1,H_2,H_3,I_4^{\rm o}).
		\label{eq:field-equality-four}
	\end{equation}
	
It remains to show that no further rational first integral can depend on the variables that evolve along the characteristics.  The change of variables $ (u_0,u_1,u_2,v_0,v_1,v_2) \longleftrightarrow (\rho,\delta,\alpha,\beta,s,u_0)$ is birational on $u_0v_0\neq0$.  In these coordinates,
	\begin{equation}
		\mathcal X = \ii\left( \partial_s+2\lambda u_0\partial_{u_0} \right).
		\label{eq:rational-proof-vector-j22}
	\end{equation}
Letting $\mathbb F := \mathbb C(\rho,\delta,\alpha,\beta) = \mathbb C(H_1,H_2,H_3,I_4^{\rm o})$ be the field generated by the four rational characteristics already identified.  Since $\mathcal X=\ii D$ in the birational coordinates $(\rho,\delta,\alpha,\beta,s,u_0)$, where $D:=\partial_s+2\lambda u_0\partial_{u_0}$, it remains to determine the kernel of $D$ in $\mathbb F(s,u_0)$. We consider therefore
	\begin{equation}
		f\in\mathbb F(s,u_0), \qquad Df=0. \label{eq:f-rational}
	\end{equation}
Choosing a reduced representation
	\begin{equation}
		f=\frac{P}{Q}, \qquad P,Q\in\mathbb F(s)[u_0], \qquad \gcd(P,Q)=1, \label{eq:PQ-rational}
	\end{equation}
the condition $Df=0$ is equivalent to
	\begin{equation}
		(DP)Q-P(DQ)=0.
		\label{DPDQ}
	\end{equation}
Since $P$ and $Q$ are coprime in $\mathbb F(s)[u_0]$, implies $ P\mid DP$, $Q\mid DQ. $ Moreover, $D$ does not increase the degree in $u_0$.  The quotients $DP/P$ and $DQ/Q$ must therefore be independent of $u_0$, so there exist $c_P(s),c_Q(s)\in\mathbb F(s)$ such that
	\begin{equation}
		DP=c_P(s)P, \qquad DQ=c_Q(s)Q.
	\end{equation}
Substitution back into \eqref{DPDQ} gives
	\begin{equation}
	c_P(s)=c_Q(s)=:c(s).
	\end{equation}
Next, writing
	\begin{equation}
		P=\sum_k p_k(s)u_0^k,
		\label{eq:P-expansion}
	\end{equation}
every nonzero coefficient satisfies
	\begin{equation}
		p_k'(s)+2\lambda k\,p_k(s)=c(s)p_k(s).
		\label{eq:pk-equation}
	\end{equation}
If two coefficients $p_k$ and $p_\ell$ with $k\neq\ell$ were both nonzero, their ratio would obey
	\begin{equation}
		\frac{\dd}{\dd s} \ln \!\left(\frac{p_k}{p_\ell}\right) = -2\lambda(k-\ell). \label{eq:log-ratio}
	\end{equation}
	The left-hand side is the logarithmic derivative of a nonzero rational
	function of $s$, and therefore cannot be a nonzero constant, since this would force an exponential dependence on $s$.  Hence
	$P$ contains only one power of $u_0$.  The same argument applies to
	$Q$, and consequently
	\begin{equation}
		f=r(s)u_0^m
		\label{eq:f-monomial}
	\end{equation}
	for some $m\in\mathbb Z$ and $r(s)\in\mathbb F(s)$.  The equation
	$Df=0$ now reduces to
	\begin{equation}
		r'(s)+2\lambda m\,r(s)=0.
		\label{eq:r-final}
	\end{equation}
	Since $r$ is rational and $\lambda\neq0$, this is possible only for
	$m=0$ and $r\in\mathbb F$.  Thus no additional nonconstant rational
	invariant arises from $(s,u_0)$, and the complete rational
	first-integral field is
	\begin{equation}
		\mathbb C(x,y,z,p_x,p_y,p_z)^{\mathcal X} = \mathbb C(H_1,H_2,H_3,I_4^{\rm o}).
		\label{eq:rational-field-short}
	\end{equation}
where the superscript $\mathcal X$ denotes the invariant field of the derivation $\mathcal X$. In particular, the rational invariant field has transcendence degree four, so no fifth functionally independent rational first integral exists.

The remaining characteristic $\zeta$ is non-rational.  Since
	\begin{equation}
		u_0=\ii(p_x-p_y)-\lambda(x+y)=: A_+ , \qquad s=-\frac{C_+}{2gA_+}, \label{eq:A-C-s}
	\end{equation}
where
	\begin{equation}
		C_+ := \ii p_z+g(x+y)-\lambda z, \label{eq:Cplus}
	\end{equation}
equation \eqref{eq:chi-def} may be rewritten as
	\begin{equation}
		I_5:=\zeta = \ln A_+ + \frac{\lambda}{g}\frac{C_+}{A_+}. \label{eq:I5-short}
	\end{equation}
This is the fifth local characteristic of the common Hamiltonian flow.
	
To determine its rôle on the real phase space, we decompose it into its real and imaginary part
	\begin{equation}
		I_5=I_5^{\rm R}+\ii I_5^{\rm I}
		\label{eq:I5-real-imag}
	\end{equation}
on a fixed logarithmic branch.  Its real part is
	\begin{equation}
		I_5^{\rm R} = \frac{1}{2}\ln \Delta + \frac{\lambda}{2g} \frac{H_1-H_2}{\Delta}, \label{eq:I5real-short}
	\end{equation}
and is therefore functionally dependent on the three Hamiltonians. The genuinely new information is contained in t$I_5^{\rm I}$, which may be written explicitly as
	\begin{equation}
			I_5^{\rm I} ={} \arg\!\left[-\lambda(x+y) +\ii(p_x-p_y)\right] -\frac{\lambda}{g} \frac{ (p_x-p_y)\left[g(x+y)-\lambda z\right] +\lambda(x+y)p_z }{ (p_x-p_y)^2+\lambda^2(x+y)^2 }.
		\label{eq:I5imag-short}
	\end{equation}
The real structure of the characteristic coordinates can also be seen directly.  On real phase space,
	\begin{equation}
		v_0=-\overline{u_0}, \qquad t=-\overline{s}, \qquad \beta=\overline{\alpha}, \label{eq:real-relations}
	\end{equation}
and hence
	\begin{equation}
		\rho=-\Delta<0, \qquad \delta=2\Re s, \qquad \Re\zeta = \frac12\ln (-\rho)-\lambda\delta. \label{eq:real-characteristics-relations}
	\end{equation}
The five independent real characteristics may therefore be chosen as
	\begin{equation}
		\rho,\qquad \delta,\qquad \Re\alpha,\qquad \Im\alpha,\qquad \Im\zeta, \label{eq:five-real-characteristics}
	\end{equation}
while the remaining real coordinate along the orbit may be taken to
	be $\Im s$, for which
	\begin{equation}
		\mathcal X(\Im s)=1.
		\label{eq:orbit-real-coordinate}
	\end{equation}
Thus every local smooth real first integral is an arbitrary smooth function of the five quantities in \eqref{eq:five-real-characteristics}.  In particular, on a generic open subset,
	\begin{equation}
		\operatorname{rank}\bigl(dH_1,dH_2,dH_3,dI_4^{\rm o},dI_5^{\rm I}\bigr)=5.
	\end{equation}

The polynomial and rational first-integral sectors therefore have generic functional rank four, whereas the transcendental characteristic raises the local smooth rank to five.

The logarithmic representative $I_5$ is branch dependent, but the fifth invariant can be globalised on a natural invariant domain.  On the real phase space,
	\begin{equation}
		|A_+|^2 = (p_x-p_y)^2+\lambda^2(x+y)^2 = \Delta. \label{eq:AplusDelta}
	\end{equation}
Hence $A_+\neq0$ is equivalent to $\Delta>0$.  Since $\Delta$ is a first integral, the open set $\mathcal U:=\{\Delta>0\}$ is invariant under the flow.  On $\mathcal U$ the branch-independent quantity
	\begin{equation}
		Z_5 = A_+\exp\!\left( \frac{\lambda}{g}\frac{C_+}{A_+} \right) \label{eq:Z5}
	\end{equation}	is globally defined, single-valued and nonvanishing.
	
On every local logarithmic branch one has $Z_5=e^{I_5}$. Consequently the real and imaginary parts of $Z_5$ carry the same additional functional information as the phase $I_5^{\rm I}$. Together with $H_1,H_2,H_3,I_4^{\rm o}$ they therefore have generic functional rank five on $\mathcal U$.  The model is thus minimally superintegrable in the polynomial and rational classes, but maximally superintegrable on the invariant open set $\Delta>0$ when transcendental first integrals are admitted.  The invariant $Z_5$ remains transcendental and does not in general admit a regular extension through $\Delta=0$.
	
\section{Weyl quantisation and quantum symmetries}
	\label{sec:quantum}
	
We now quantise the three Poisson descriptions, beginning with their polynomial first integrals.  We employ Weyl quantisation throughout; for standard accounts of the Weyl calculus and its phase-space formulation, see \cite{groenewold1946pr,moyal1949qu}. In the polynomial sector the relations below are identities of differential expressions on the Schwartz space $\mathcal S(\mathbb R^3)$ of smooth functions whose derivatives decay faster than any inverse power at infinity. Real Weyl symbols define formally symmetric operators on this common invariant space.  We make no assertion of self-adjointness or strong commutation.

Using the block representation of the Poisson tensors $J_r$ introduced in equation \eqref{Jblock}, we realise the original phase-space variables in terms of canonical Darboux variables $(\boldsymbol Q,\boldsymbol P)$. We then define
	\begin{equation}
		\widehat{\boldsymbol Q}=(x,y,z)^{\mathsf T}, \qquad \widehat{\boldsymbol P} =-\ii\hbar(\partial_x,\partial_y,\partial_z)^{\mathsf T},
	\end{equation}
so that
	\begin{equation}
		[\widehat Q_i,\widehat P_j] =\ii\hbar\delta_{ij},
	\end{equation}
with $\hbar$ denoting the reduced Planck constant. A realisation of the original variables associated with $J_r$ is then
	\begin{equation}
		\widehat{\boldsymbol q}^{(r)} = \widehat{\boldsymbol Q}, \qquad \widehat{\boldsymbol p}^{(r)} = M_r^{\mathsf T}\widehat{\boldsymbol P}.
		\label{eq:Darboux-short}
	\end{equation}
	Consequently, we have the commutation relations
	\begin{equation}
		[\widehat q_i^{(r)},\widehat p_j^{(r)}] = \ii\hbar(M_r)_{ij},
		\qquad
		[\widehat q_i^{(r)},\widehat q_j^{(r)}] = [\widehat p_i^{(r)},\widehat p_j^{(r)}] =0.
		\label{eq:noncanonical-CCR}
	\end{equation}
	Explicitly, in each representation we identify
	\begin{equation}
		\begin{array}{c|ccc}
			r &
			\widehat p_x^{(r)}&
			\widehat p_y^{(r)}&
			\widehat p_z^{(r)}
			\\ \hline
			1&
			\widehat P_x&
			\widehat P_y&
			\widehat P_z
			\\
			2&
			2\widehat P_x-\widehat P_y+\widehat P_z&
			\widehat P_x+\widehat P_z&
			\widehat P_x-\widehat P_y+\widehat P_z
			\\
			3&
			2\widehat P_x-\widehat P_y&
			\widehat P_x&
			\widehat P_z
		\end{array}.
		\label{eq:momentum-table}
	\end{equation}
A useful consequence is the realisation-independent identity
	\begin{equation}
		\widehat p_x^{(r)}-\widehat p_y^{(r)} = -\ii\hbar(\partial_x-\partial_y), \qquad r=1,2,3.
		\label{eq:universal-pdiff-short}
	\end{equation}
	
For a classical polynomial we first make the substitution $\boldsymbol p\mapsto M_r^{\mathsf T}\boldsymbol P$ and then apply symmetric Weyl ordering with respect to the canonical pairs $(x,P_x)$, $(y,P_y)$ and $(z,P_z)$.  Thus, for example, $xP_x$ is represented by $(x\widehat P_x+\widehat P_xx)/2$.
	
Writing the quadratic Hamiltonians as
	\begin{equation}
		H_a = \boldsymbol p^{\mathsf T}G_a\boldsymbol p + V_a(\boldsymbol q), \qquad G_a = \frac12 \partial_{\boldsymbol p} \partial_{\boldsymbol p}^{\mathsf T}H_a,
		\label{eq:H-GV}
	\end{equation}
their nine quantum realisations take the compact form
	\begin{equation}
		\widehat H_a^{(r)} = -\hbar^2 \nabla^{\mathsf T} (M_rG_aM_r^{\mathsf T}) \nabla + V_a(x,y,z), \qquad a,r=1,2,3.
		\label{eq:all-nine-H-short}
	\end{equation}
A direct calculation gives
	\begin{equation}
		[\widehat H_a^{(r)},\widehat H_b^{(r)}] =0, \qquad a,b,r=1,2,3,
		\label{eq:qHcomm-short}
	\end{equation}
so the three quadratic Hamiltonians remain mutually commuting in every realisation.
	
We next quantise the momentum-odd degree-four integral $I_4^{\rm o}$.  It is convenient to use the four quadratic blocks introduced in section 2.  The realisation-independent operators are
	\begin{equation}
		\widehat A = -\hbar^2(\partial_x-\partial_y)^2 -\lambda^2(x+y)^2, \qquad \widehat D = -\ii\hbar(x+y)(\partial_x-\partial_y),
		\label{eq:ADhat-j22}
	\end{equation}
	while Weyl ordering the remaining two blocks gives
	\begin{equation}
		\begin{aligned}
			\widehat B^{(1)}
			={}& \ii\hbar\Big[ -(gz+\lambda y)(\partial_x-\partial_y) +(g(x+y)-\lambda z)\partial_z +\lambda(x+y)\partial_y +\frac{\lambda}{2} \Big], \\ 
			\widehat C^{(1)}={}& \hbar^2 (2\partial_x\partial_y-2\partial_y^2-\partial_z^2) -2\lambda(x+y)(gz+\lambda y) -\bigl[g(x+y)-\lambda z\bigr]^2, \\
			\widehat B^{(2)} ={}& \ii\hbar\Big[ ((g+\lambda)x+gy-(g+\lambda)z)(\partial_x-\partial_y) 
			+((g+\lambda)(x+y)-\lambda z)\partial_z +\lambda(x+y)\partial_y +\frac{\lambda}{2} \Big], \\ \widehat C^{(2)}
			={}& \hbar^2(\partial_x^2-\partial_y^2-\partial_z^2) -2\lambda(x+y)(gz+\lambda y) -\bigl[g(x+y)-\lambda z\bigr]^2, \\[1ex] \widehat B^{(3)}
			={}& \ii\hbar\Big[ (\lambda x-gz)(\partial_x-\partial_y) +(g(x+y)-\lambda z)\partial_z +\lambda(x+y)\partial_y +\frac{\lambda}{2} \Big], \\ \widehat C^{(3)}
			={}& \hbar^2 (2\partial_x^2-2\partial_x\partial_y-\partial_z^2) -2\lambda(x+y)(gz+\lambda y) -\bigl[g(x+y)-\lambda z\bigr]^2.
		\end{aligned}
		\label{eq:BChat-j22}
	\end{equation}
	The Weyl image of the complete classical polynomial is therefore
	\begin{equation}
		\widehat I_4^{\rm o\,(r)} = 128g^6 \left[ \frac12 \{\widehat A,\widehat B^{(r)}\}_{+} - \frac{\lambda}{2} \{\widehat D,\widehat C^{(r)}\}_{+} \right],
		\label{eq:I4blocks-j22}
	\end{equation}
where we use the anticommutator $\{A,B\}_{+}:=AB+BA$. Although the classical polynomial has total phase-space degree four, the operator $\widehat I_4^{\rm o\,(r)}$ has differential order three. The corresponding fully expanded third-order differential operators are collected in Appendix~\ref{app:I4-explicit}.
	
The same operator also retains the compact hidden-algebra factorisation obtained classically.  If $\widehat h^{(r)},\widehat e^{(r)}, \widehat v_1^{(r)},\widehat v_2^{(r)}$ denote the Weyl quantisations of the corresponding classical symbols introduced in section 3, then
	\begin{equation}
		\widehat I_4^{\rm o\,(r)} = 32\ii g^6\lambda \left[ \{\widehat h^{(r)},\widehat v_2^{(r)}\}_{+} + \frac12 \{\widehat e^{(r)},\widehat v_1^{(r)}\}_{+} - \frac{g}{\lambda} \{\widehat e^{(r)},\widehat v_2^{(r)}\}_{+} \right].
		\label{eq:I4hiddenquantum-short}
	\end{equation}
For $r=1$ this expression involves the genuine hidden $\mathfrak u(2,1)$ generators of the canonical realisation.  For $r=2,3$ the same notation denotes the Weyl quantisations of the same classical phase-space functions in the corresponding realisations. The complete nine-generator family does not reproduce the same $\mathfrak u(2,1)$ commutation relations.  This distinction does not affect their use in the compact representation \eqref{eq:I4hiddenquantum-short}.

The commutator table is the direct quantum counterpart of the classical table in section 2:
	\begin{equation}
		[\widehat H_a^{(r)},\widehat I_4^{\rm o\,(r)}] = \ii\hbar c_{ra}\widehat K, \qquad a,r=1,2,3,
		\label{eq:qI4table-short}
	\end{equation}
where $c_{ra}$ are the coefficients listed in Table~\ref{tab:classical-table}.  The operator
	\begin{equation}
		\widehat\Delta := \widehat H_1^{(r)}-\widehat H_3^{(r)} = -\hbar^2(\partial_x-\partial_y)^2 +\lambda^2(x+y)^2
		\label{eq:Deltahat}
	\end{equation}
is independent of $r$, and we write
	\begin{equation}
		\widehat K := 256g^8\widehat\Delta^{\,2}.
		\label{eq:Khat}
	\end{equation}
In particular, we find
	\begin{equation}
		[\widehat H_r^{(r)},\widehat I_4^{\rm o\,(r)}] =0, \qquad r=1,2,3,
		\label{eq:I4o-conservation-quantum}
	\end{equation}
so $\widehat I_4^{\rm o\,(r)}$ is an exact third-order differential symmetry of each paired Hamiltonian. 
Under the invertible Darboux transformation, the classical symbols of $\widehat H_1^{(r)},\widehat H_2^{(r)},\widehat H_3^{(r)}$ and $\widehat I_4^{\rm o\,(r)}$ correspond to the four generically independent functions $H_1,H_2,H_3,I_4^{\rm o}$ identified in Section~2.  Thus, in the usual symbol-level sense, $\widehat I_4^{\rm o\,(r)}$ supplies the additional quantum symmetry required for superintegrability.

 Moreover, we compute
	\begin{equation}
		[\widehat\Delta,\widehat I_4^{\rm o\,(r)}] = \frac{\ii\hbar\lambda}{g^2}\widehat K, \qquad [\widehat K,\widehat I_4^{\rm o\,(r)}] = 512\ii\hbar\lambda g^6 \widehat\Delta\,\widehat K.
		\label{eq:IKclosure-j22}
	\end{equation}
	
We next quantise the momentum-even degree-four integral $I_4^{\rm e}=F^2+G^2$.  Let $\widehat F^{(r)}$ and $\widehat G^{(r)}$ denote the Weyl images of the quadratic functions $F$ and $G$ introduced in section 4.  In terms of the realisation-dependent momentum operators they are
	\begin{equation}
		\begin{aligned}
			\widehat F^{(r)} ={}& (\widehat p_x^{(r)})^2 -(\widehat p_y^{(r)})^2 -(\widehat p_z^{(r)})^2 +g^2(x+y)^2 -\lambda^2x^2+\lambda^2y^2+\lambda^2z^2, \\
			\widehat G^{(r)} ={}& -g\{z,\widehat p_x^{(r)}-\widehat p_y^{(r)}\}_{+} +g\{x+y,\widehat p_z^{(r)}\}_{+} +\lambda \left( \{x,\widehat p_x^{(r)}\}_{+} +\{y,\widehat p_y^{(r)}\}_{+} -\{z,\widehat p_z^{(r)}\}_{+} \right).
		\end{aligned}
		\label{eq:FGhat-short}
	\end{equation}
Weyl ordering the complete classical polynomial $F^2+G^2$ gives
	\begin{equation}
		\widehat I_4^{\rm e\,(r)} = (\widehat F^{(r)})^2 + (\widehat G^{(r)})^2 - 6\hbar^2\lambda^2.
		\label{eq:I4ehat-short}
	\end{equation}
The scalar term is required because \eqref{eq:I4ehat-short} is the Weyl image of the specified classical polynomial.  It is not required merely to make the resulting operator commute with the paired Hamiltonian.

The quantum counterpart of the classical bracket relation is
	\begin{equation}
		[\widehat H_a^{(r)},\widehat I_4^{\rm e\,(r)}] = \ii\hbar c_{ra} \frac{\widehat I_4^{\rm o\,(r)}}{8g^4}, \qquad a,r=1,2,3.
		\label{eq:qI4etable-short}
	\end{equation}
Hence we have
	\begin{equation}
		[\widehat H_r^{(r)},\widehat I_4^{\rm e\,(r)}] =0, \qquad r=1,2,3,
		\label{eq:I4e-conservation-quantum}
	\end{equation}
and $\widehat I_4^{\rm e\,(r)}$ is an exact fourth-order differential symmetry.

The commutator of the two higher-order symmetries also closes.  Using $\Delta$, $\Theta$ and $L_r$ as defined in section 4, we find
	\begin{equation}
		[\widehat I_4^{\rm o\,(r)}, \widehat I_4^{\rm e\,(r)}] = 256\ii\hbar g^6 \left[ 2\lambda \widehat{\Delta I_4^{\rm e}}^{(r)} - \widehat{\Theta L_r}^{(r)} + 12\hbar^2\lambda^3\widehat\Delta \right].
		\label{eq:qI4I4-short}
	\end{equation}
Here and below, a hat over a composite classical expression denotes the Weyl quantisation of the complete expression in the $r$th realisation, rather than the ordinary product of the separately quantised factors. The last term in \eqref{eq:qI4I4-short} has no classical counterpart and is a genuine quantum ordering contribution.
	
The same distinction is important for the classical octic relation. Weyl ordering the complete classical identity gives
	\begin{equation}
		4\widehat{\Delta^2 I_4^{\rm e}}^{(r)} - \widehat{\Theta^2}^{(r)} - \frac{1}{1024g^{12}} \widehat{(I_4^{\rm o})^2}^{(r)} =0.
		\label{eq:q-octic-short}
	\end{equation}
Equation \eqref{eq:q-octic-short} should not be interpreted as an ordinary-product identity among the separately quantised generators. In general,
	\begin{equation}
		4\widehat\Delta^{\,2} \widehat I_4^{\rm e\,(r)} - (\widehat\Theta^{(r)})^2 - \frac{ (\widehat I_4^{\rm o\,(r)})^2 }{1024g^{12}} \neq0.
		\label{eq:ordinary-product-nonzero}
	\end{equation}
Already in the canonical realisation the residual is a nonzero differential operator of order five.  Thus the Weyl image of the classical octic relation is not an ordinary polynomial identity among the separately quantised generators.

The finite-order differential symmetries constructed above establish quantum superintegrability for each of the three paired Hamiltonian realisations.  The classical analysis of section~\ref{sec:pde-completion} shows, however, that the fifth independent classical invariant required for maximal superintegrability lies outside the rational class.  We now show that its branch-free transcendental representative $Z_5$ admits a natural nonlocal quantum counterpart.

We introduce the first-order operators
	\begin{equation}
		\widehat A_+^{(r)} = \ii\bigl( \widehat p_x^{(r)}-\widehat p_y^{(r)} \bigr) -\lambda(x+y), \qquad \widehat C_+^{(r)} = \ii\widehat p_z^{(r)} +g(x+y)-\lambda z .
		\label{eq:AplusCplus-quantum}
	\end{equation}
The realisation-independent identity \eqref{eq:universal-pdiff-short} implies
	\begin{equation}
		\widehat A_+^{(r)} = \hbar(\partial_x-\partial_y)-\lambda(x+y), \qquad r=1,2,3.
		\label{eq:Aplus-universal}
	\end{equation}
Moreover,
	\begin{equation}
		[\widehat A_+^{(r)},\widehat C_+^{(r)}]=0, \qquad r=1,2,3, \label{eq:ACcommute}
	\end{equation}
so no ordering ambiguity arises in the quantum analogue of $C_+/A_+$.
	
The paired Hamiltonians satisfy the exact intertwining relations
	\begin{equation}
		[\widehat H_r^{(r)},\widehat A_+^{(r)}] = 2\hbar\lambda\,\widehat A_+^{(r)}, \qquad [\widehat H_r^{(r)},\widehat C_+^{(r)}] = 2\hbar\lambda\,\widehat C_+^{(r)} - 2\hbar g\,\widehat A_+^{(r)}.
		\label{eq:AC-intertwining}
	\end{equation}
Whenever $(\widehat A_+^{(r)})^{-1}$ is defined on a common	invariant domain, it follows that
	\begin{equation}
		[\widehat H_r^{(r)}, (\widehat A_+^{(r)})^{-1}] = -2\hbar\lambda (\widehat A_+^{(r)})^{-1}.
		\label{eq:Ainverse-comm}
	\end{equation}
Hence, defining
	\begin{equation}
		\widehat \Xi^{(r)} := \frac{\lambda}{g} \widehat C_+^{(r)} (\widehat A_+^{(r)})^{-1},
		\label{eq:B-nonlocal}
	\end{equation}
we obtain
	\begin{equation}
		[\widehat H_r^{(r)},\widehat \Xi^{(r)}] = -2\hbar\lambda\,\mathbf 1 .
		\label{eq:B-intertwining}
	\end{equation}
Since the right-hand side is central, it follows formally, and on any common invariant domain of analytic vectors on which the exponential is defined, that
	\begin{equation}
		[\widehat H_r^{(r)},e^{\widehat \Xi^{(r)}}] = -2\hbar\lambda\,e^{\widehat \Xi^{(r)}}.
		\label{eq:expB-intertwining}
	\end{equation}
Combining this with the first relation in \eqref{eq:AC-intertwining} gives the branch-free nonlocal symmetry
	\begin{equation}
		\widehat Z_5^{(r)} := \widehat A_+^{(r)} \exp\!\left[ \frac{\lambda}{g} \widehat C_+^{(r)} (\widehat A_+^{(r)})^{-1} \right],
		\label{eq:Z5hat}
	\end{equation}
for which
	\begin{equation}
		[\widehat H_r^{(r)},\widehat Z_5^{(r)}]=0
		\label{eq:Z5hat-comm}
	\end{equation}
on every common invariant domain on which the operators entering \eqref{eq:Z5hat} are defined.

The construction of $\widehat Z_5^{(r)}$ involves $(\widehat A_+)^{-1}$, so it remains to determine whether this inverse can be defined as an operator rather than merely formally.  This becomes transparent after a partial Fourier transformation in the variable $v$.  Indeed, in the coordinates
\begin{equation}
	u=x+y,\qquad v=x-y,
\end{equation}
one has
\begin{equation}
\widehat A_+=2\hbar\partial_v-\lambda u.
\end{equation}
The partial Fourier transform in $v$ is unitary on $L^2$ and sends $\partial_v$ to multiplication by $\ii k$.  Hence $\widehat A_+$ is unitarily equivalent to multiplication by
\begin{equation}
	a(u,k)=-\lambda u+2\ii\hbar k.
	\label{eq:a-uk}
\end{equation}
This representation makes both the kernel and the inverse explicit. For real $u$ and $k$, $a(u,k)=0$ only when $u=k=0$, which is a set of measure zero.  Therefore the corresponding multiplication operator has trivial $L^2$ kernel.  Its inverse is multiplication by $a(u,k)^{-1}$ on the domain
\begin{equation}
\mathcal D(\widehat A_+^{-1}) = \left\{ \psi\in L^2: a^{-1}\psi\in L^2 \right\}.
\end{equation}
This domain is dense, although the inverse is unbounded because $a(u,k)^{-1}$ becomes singular as $(u,k)\to(0,0)$.

The exponential in \eqref{eq:Z5hat} can be treated on the same representation.  There we can write
	\begin{equation}
		\widehat C_+^{(r)} = \hbar\partial_z-\lambda z+\gamma_r(u,k),
	\end{equation}
with
	\begin{equation}
		\gamma_1=\gamma_3=gu, \qquad \gamma_2=gu+2\ii\hbar k.
	\end{equation}
Hence the exponent in \eqref{eq:Z5hat} is a first-order affine differential operator in $z$. Its exponential is well defined on a natural dense class of analytic vectors whose partial Fourier support is bounded away from $a(u,k)=0$.  On this class $(\widehat A_+^{(r)})^{-1}$ and the exponential entering \eqref{eq:Z5hat} are both well defined, so $\widehat Z_5^{(r)}$ is realised as a genuine densely defined nonlocal operator rather than merely as a formal conserved expression.

This branch-free construction is preferable to quantising the logarithmic characteristic $I_5$ directly, since it avoids introducing an operator logarithm and its associated branch choice.  The polynomial symmetry $\widehat I_4^{\rm o\,(r)}$, together with the three quadratic symmetries, establishes quantum superintegrability in the symbol-level sense described above, while $\widehat I_4^{\rm e\,(r)}$ provides an additional fourth-order polynomial symmetry.  The operator $\widehat Z_5^{(r)}$ gives a nonlocal quantum counterpart of the fifth classical invariant.  A complete operator-theoretic analysis, including its maximal domain, adjoint properties and the appropriate notion of independence from the polynomial symmetry algebra, is beyond the scope of the present work.

\section{Conclusions}

We have analysed the classical and quantum superintegrable structure of a resonant three-dimensional ghostly Hamiltonian admitting three Poisson descriptions of the same flow.  The three quadratic Hamiltonians $H_1,H_2,H_3$ are mutually in involution with respect to each Poisson tensor, while the momentum-odd degree-four integral $I_4^{\rm o}$ is conserved by every paired system $(H_r,J_r)$ and is generically independent of the three quadratic Hamiltonians.  This establishes classical superintegrability for all three Hamiltonian descriptions.  In the canonical realisation, $I_4^{\rm o}$ also admits a compact factorisation in terms of the hidden spectrum-generating $\mathfrak u(2,1)$ algebra.

We found that the Jordan-chain structure of the common flow provides a systematic classification of the polynomial first integrals.  Besides $I_4^{\rm o}$, there is a second primitive degree-four generator $I_4^{\rm e}$ of even momentum parity.  Using the Jordan-Chevalley decomposition of the derivation and the invariant theory of its locally nilpotent part, we showed that the complete polynomial first-integral ring is generated by $H_1,H_2,H_3,I_4^{\rm o},I_4^{\rm e}$, subject to a single octic relation.  Consequently the polynomial ring has transcendence degree four, so $I_4^{\rm e}$ is polynomially primitive without providing a fifth functionally independent integral.

Going beyond the polynomial class, we solved the complete first-integral equation in coordinates adapted to the two Jordan chains.  This yields a complete local characteristic description and determines the rational invariant field, $ \mathbb C(\text{rational first integrals}) = \mathbb C(H_1,H_2,H_3,I_4^{\rm o})$. Thus no fifth independent rational first integral exists, irrespective of momentum order.  The missing functional degree of freedom is instead supplied by a transcendental characteristic.  The transcendental sector raises the generic smooth functional rank to five on the invariant open set $\Delta>0$.  The system is therefore minimally superintegrable in the polynomial and rational classes, but maximally superintegrable on $\Delta>0$ when transcendental first integrals are admitted.  A branch-independent representative is provided by the exponential invariant $Z_5$, although the singular set $\Delta=0$ prevents the present construction from yielding a regular extension
across this boundary.

At the quantum level, the three Poisson structures were treated uniformly by passing to Darboux coordinates and applying Weyl quantisation.  The polynomial integrals give exact higher-order differential symmetries $\widehat I_4^{\rm o\,(r)}$ and $\widehat I_4^{\rm e\,(r)}$ of differential orders three and four, respectively.  Their commutator algebra reproduces the classical structure together with genuine quantum ordering contributions.  In particular, the Weyl image of the classical octic relation is not an ordinary polynomial identity among the separately quantised generators.

The branch-free transcendental invariant also admits the nonlocal quantum counterpart $\widehat Z_5^{(r)}$ constructed in section~\ref{sec:quantum}.  Its commutator with the corresponding Hamiltonian vanishes on a suitable common invariant domain.  A partial Fourier transformation shows that the first-order operator entering its inverse is unitarily equivalent to a multiplication operator with trivial $L^2$ kernel, so the inverse is densely defined, although unbounded.  The exponential can likewise be defined on a suitable class of analytic vectors.  The resulting nonlocal symmetry therefore goes beyond a purely formal analogue of the classical transcendental invariant.

Several analytic questions remain open.  In particular, a complete operator-theoretic treatment of $\widehat Z_5^{(r)}$ would require a systematic analysis of its maximal invariant domain, adjoint and spectral properties, together with an appropriate notion of independence from the polynomial symmetry algebra.  These questions are needed before a precise statement of quantum maximal superintegrability can be made.  More broadly, the present model exhibits a hierarchy of invariant classes in which the polynomial and rational sectors both have generic functional rank four, while maximal superintegrability is achieved only after a transcendental completion. It would be interesting to determine how generally this mechanism occurs in higher-dimensional resonant Jordan-chain systems and other non-semisimple Hamiltonian models.

\begin{appendices}
		
		\numberwithin{equation}{section}
		\renewcommand{\theequation}{\thesection\arabic{equation}}
		
\section{Hidden $\mathfrak u(2,1)$ structure and noncanonical realisations}
	\label{app:hidden-symbols}
	
	For completeness, we collect here the remaining classical symbols and Poisson-algebra relations underlying the hidden $\mathfrak u(2,1)$ description used in Section~\ref{sec:hidden-form}.  These details are not required for the functional-rank arguments in the main text, but they make the connection with the spectrum-generating algebra of Ref.~\cite{FringMarquette2026} explicit.
	
	The $\mathfrak{sl}_2$ lowering symbol complementary to
	$h_{\mathrm{cl}}$ and $e_{\mathrm{cl}}$ is
	\begin{equation}
		f_{\mathrm{cl}}
		=
		2\ii\left[
		\frac{g^2}{\lambda^2}z(p_y-p_x)
		+\frac{g}{2\lambda}
		\bigl((x-y)(p_x-p_y)-2zp_z\bigr)
		+\frac12
		\bigl(p_z(x-y)-z(p_x+p_y)\bigr)
		\right].
		\label{eq:fcl-j22}
	\end{equation}
	With respect to the canonical Poisson tensor $J_1$,
	\begin{equation}
		\{h_{\mathrm{cl}},e_{\mathrm{cl}}\}_{J_1}
		=
		-2\ii e_{\mathrm{cl}},
		\qquad
		\{h_{\mathrm{cl}},f_{\mathrm{cl}}\}_{J_1}
		=
		2\ii f_{\mathrm{cl}},
		\qquad
		\{e_{\mathrm{cl}},f_{\mathrm{cl}}\}_{J_1}
		=
		-2\ii h_{\mathrm{cl}}.
		\label{eq:sl2-j22}
	\end{equation}
	
The five symbols $(v_m)_{\mathrm{cl}}$, $m=-2,-1,0,1,2$, form the corresponding spin-two multiplet.  In addition to $(v_1)_{\mathrm{cl}}$ and $(v_2)_{\mathrm{cl}}$ given in Section~\ref{sec:hidden-form}, the remaining symbols are
	\begin{align}
		(v_{-2})_{\mathrm{cl}}
		&=
		-\frac14
		\left[
		(2gz+\lambda(y-x))^2+(p_x+p_y)^2
		\right],
		\notag
		\\
		(v_{-1})_{\mathrm{cl}}
		&=
		\frac{g}{\lambda}(p_y^2-p_x^2)
		+2g\lambda z^2
		-p_z(p_x+p_y)
		+\lambda^2z(y-x),
		\notag
		\\
		(v_0)_{\mathrm{cl}}
		&=
		-2\frac{g^2}{\lambda^2}(p_x-p_y)^2
		+4\frac{g}{\lambda}p_z(p_y-p_x)
		-2g\lambda z(x+y)
		\notag
		\\
		&\qquad
		+p_x^2-p_y^2-2p_z^2
		+\lambda^2(x^2-y^2-2z^2).
		\label{eq:vm-extra-app-j22}
	\end{align}
	Their weights with respect to $h_{\mathrm{cl}}$ are
	\begin{equation}
		\{h_{\mathrm{cl}},(v_m)_{\mathrm{cl}}\}_{J_1}
		=
		-2\ii m(v_m)_{\mathrm{cl}},
		\qquad
		m=-2,-1,0,1,2.
		\label{eq:vmweights-app-j22}
	\end{equation}
The generators $e_{\mathrm{cl}}$ and $f_{\mathrm{cl}}$ raise and lower the index $m$, respectively, with
	\begin{equation}
		\{e_{\mathrm{cl}},(v_2)_{\mathrm{cl}}\}_{J_1}=0,
		\qquad
		\{f_{\mathrm{cl}},(v_{-2})_{\mathrm{cl}}\}_{J_1}=0.
		\label{eq:highest-lowest-app}
	\end{equation}
Together with the central symbol $(K_{\mathrm{hid}})_{\mathrm{cl}}$ defined in eq.~\eqref{eq:Khid-short}, these symbols form the classical hidden-$\mathfrak u(2,1)$ realisation of the canonical system.

It is important to distinguish this canonical hidden algebra from the use of the same phase-space functions with the other two Poisson tensors.  If the symbols $h_{\mathrm{cl}},e_{\mathrm{cl}},f_{\mathrm{cl}}, (v_m)_{\mathrm{cl}}$ and $(K_{\mathrm{hid}})_{\mathrm{cl}}$ are left unchanged while the Poisson bracket is replaced by that associated with $J_2$ or $J_3$, the complete $\mathfrak u(2,1)$ table is no longer preserved.  For example,
	\begin{align}
		\{h_{\mathrm{cl}},(v_1)_{\mathrm{cl}}\}_{J_1}
		&=
		-2\ii(v_1)_{\mathrm{cl}},
		&
		\{h_{\mathrm{cl}},(v_1)_{\mathrm{cl}}\}_{J_2}
		&=
		-2\ii(v_1)_{\mathrm{cl}}
		-4\ii(v_2)_{\mathrm{cl}},
		\notag
		\\
		\{(K_{\mathrm{hid}})_{\mathrm{cl}},h_{\mathrm{cl}}\}_{J_1}
		&=0,
		&
		\{(K_{\mathrm{hid}})_{\mathrm{cl}},h_{\mathrm{cl}}\}_{J_3}
		&=
		4\ii\frac{\lambda}{g}(v_2)_{\mathrm{cl}}.
		\label{eq:hidden-deform-j22}
	\end{align}
Thus the same-symbol realisations associated with $J_2$ and $J_3$ should not be interpreted as two further identical copies of the hidden $\mathfrak u(2,1)$ algebra.

An algebra-preserving realisation can instead be obtained by transporting the canonical symbols with the linear transformation relating $J_r$ to $J_1$.  If $X_{\mathrm{cl}}(q,p)$ denotes any canonical hidden-algebra symbol, define
	\begin{equation}
		\widetilde X_{\mathrm{cl}}^{(r)}(q,p)
		=
		X_{\mathrm{cl}}
		\left(
		q,M_r^{-\mathsf T}p
		\right),
		\label{eq:hidden-pullback-j22}
	\end{equation}
where the matrices $M_r$ are given in \eqref{eq:Mmatrices-short}.  For any two canonical symbols $X_{\mathrm{cl}}$ and $Y_{\mathrm{cl}}$, this transformation satisfies
	\begin{equation}
		\left\{ \widetilde X_{\mathrm{cl}}^{(r)}, \widetilde Y_{\mathrm{cl}}^{(r)} \right\}_{J_r} = \widetilde{ \{X_{\mathrm{cl}},Y_{\mathrm{cl}}\}_{J_1} }^{(r)} .
		\label{eq:pullback-poisson}
	\end{equation}
Hence the pulled-back symbols satisfy the same $\mathfrak u(2,1)$ Poisson relations with respect to $J_r$ as the original symbols do with respect to $J_1$.

There is, however, a complementary disadvantage to this algebra-adapted description.  For generic $g$ and $\lambda$, the three Hamiltonians no longer lie in the linear span of $1$ and the nine pulled-back hidden generators.  Thus there are two natural but inequivalent extensions of the canonical hidden-algebra description to the noncanonical Poisson tensors: retaining the original phase-space symbols preserves the simple expressions for the Hamiltonians and $I_4^{\rm o}$ used in the main text, whereas the pullback \eqref{eq:hidden-pullback-j22} preserves the complete $\mathfrak u(2,1)$ Poisson algebra.

The present work uses the former description, since its purpose is the construction and classification of first integrals rather than a representation-theoretic analysis of the noncanonical realisations.

\section{Explicit third-order symmetry operators}
	\label{app:I4-explicit}
	
The Weyl quantisations of the momentum-odd integral $I_4^{\rm o}$ in the three Poisson realisations have the following differential expressions.  We write
	\begin{equation}
		\widehat I_4^{\rm o\,(r)} = 128\ii g^6\hbar\,\mathcal L_r,
		\label{eq:I-Lr-app-j22}
	\end{equation}
with all coordinate coefficients placed to the left of the derivatives.  Then
	\begin{align}
		\mathcal L_1={}&\hbar^2\Bigl[
		(gz+\lambda y)\partial_x^3
		-(3gz+2\lambda y-\lambda x)\partial_x^2\partial_y
		-(g(x+y)-\lambda z)\partial_x^2\partial_z
		-\frac{\lambda}{2}\partial_x^2\notag\\
		&+(3gz+\lambda y-2\lambda x)\partial_x\partial_y^2
		+2(g(x+y)-\lambda z)\partial_x\partial_y\partial_z
		+\lambda\partial_x\partial_y
		-\lambda(x+y)\partial_x\partial_z^2\notag\\
		&-(gz-\lambda x)\partial_y^3
		-(g(x+y)-\lambda z)\partial_y^2\partial_z
		-\frac{\lambda}{2}\partial_y^2
		+\lambda(x+y)\partial_y\partial_z^2\Bigr]\notag\\
		&-\lambda(x+y)\Bigl[
		\bigl([g(x+y)-\lambda z]^2+\lambda(x+y)(gz+\lambda y)\bigr)\partial_x\notag\\
		&\hspace{17mm}+\bigl(\lambda^2(x+y)^2-[g(x+y)-\lambda z]^2-\lambda(x+y)(gz+\lambda y)\bigr)\partial_y\notag\\
		&\hspace{17mm}+\lambda(x+y)[g(x+y)-\lambda z]\partial_z+\frac{\lambda^2}{2}(x+y)\Bigr],
		\label{eq:L1-app-j22}
	\end{align}
	\begin{align}
		\mathcal L_2={}&\hbar^2\Bigl[
		(-gx+(\lambda-g)y+(g+\lambda)z)\partial_x^3\notag\\
		&+(3gx+3gy-3gz+\lambda x-2\lambda y-3\lambda z)\partial_x^2\partial_y
		-((g+\lambda)(x+y)-\lambda z)\partial_x^2\partial_z
		-\frac{\lambda}{2}\partial_x^2\notag\\
		&+(-3gx-3gy+3gz-2\lambda x+\lambda y+3\lambda z)\partial_x\partial_y^2\notag\\
		&+2((g+\lambda)(x+y)-\lambda z)\partial_x\partial_y\partial_z
		+\lambda\partial_x\partial_y-\lambda(x+y)\partial_x\partial_z^2\notag\\
		&+(gx+gy-gz+\lambda x-\lambda z)\partial_y^3
		-((g+\lambda)(x+y)-\lambda z)\partial_y^2\partial_z
		-\frac{\lambda}{2}\partial_y^2
		+\lambda(x+y)\partial_y\partial_z^2\Bigr]\notag\\
		&-\lambda(x+y)\Bigl[
		\bigl([g(x+y)-\lambda z]^2+\lambda(x+y)(gz+\lambda y)\notag\\
		&\hspace{25mm}+\lambda(x+y)[g(x+y)-\lambda z]+\lambda^2(x+y)^2\bigr)\partial_x\notag\\
		&\hspace{17mm}-\bigl([g(x+y)-\lambda z]^2+\lambda(x+y)(gz+\lambda y)\notag\\
		&\hspace{35mm}+\lambda(x+y)[g(x+y)-\lambda z]\bigr)\partial_y\notag\\
		&\hspace{17mm}+\lambda(x+y)((g+\lambda)(x+y)-\lambda z)\partial_z+\frac{\lambda^2}{2}(x+y)\Bigr],
		\label{eq:L2-app-j22}
	\end{align}
	\begin{align}
		\mathcal L_3={}&\hbar^2\Bigl[
		(gz+\lambda x+2\lambda y)\partial_x^3
		-(3gz+2\lambda x+5\lambda y)\partial_x^2\partial_y
		-(g(x+y)-\lambda z)\partial_x^2\partial_z
		-\frac{\lambda}{2}\partial_x^2\notag\\
		&+(3gz+\lambda x+4\lambda y)\partial_x\partial_y^2
		+2(g(x+y)-\lambda z)\partial_x\partial_y\partial_z
		+\lambda\partial_x\partial_y
		-\lambda(x+y)\partial_x\partial_z^2\notag\\
		&-(gz+\lambda y)\partial_y^3
		-(g(x+y)-\lambda z)\partial_y^2\partial_z
		-\frac{\lambda}{2}\partial_y^2
		+\lambda(x+y)\partial_y\partial_z^2\Bigr]\notag\\
		&-\lambda(x+y)\Bigl[
		\bigl([g(x+y)-\lambda z]^2+\lambda(x+y)(gz+\lambda y)+\lambda^2(x+y)^2\bigr)\partial_x\notag\\
		&\hspace{17mm}-\bigl([g(x+y)-\lambda z]^2+\lambda(x+y)(gz+\lambda y)\bigr)\partial_y\notag\\
		&\hspace{17mm}+\lambda(x+y)[g(x+y)-\lambda z]\partial_z+\frac{\lambda^2}{2}(x+y)\Bigr].
		\label{eq:L3-app-j22}
	\end{align}
In these expressions, every derivative acts on the wavefunction to its right.  They agree with the compact block representation \eqref{eq:I4blocks-j22} and provide explicit forms suitable for direct symbolic or spectral calculations.
		
		\end{appendices}
	
\newif\ifabfull\abfulltrue

\end{document}